\documentclass[journal,12pt,onecolumn,draftclsnofoot,]{IEEEtran}
\usepackage[cmex10]{amsmath,mathtools} 
\usepackage{mathtools}
\makeatletter
\newcommand{\svast}{\bBigg@{3}}
\newcommand{\vast}{\bBigg@{4}}
\newcommand{\Vast}{\bBigg@{5}}

\makeatother

\usepackage{amsthm, amssymb, amsfonts} 
\usepackage{bm} 
\usepackage[ruled]{algorithm2e} 
\usepackage{cite} 
\usepackage[pdftex]{graphicx} 
\usepackage{epstopdf} 
\usepackage{color}
\usepackage[keeplastbox]{flushend}
\usepackage{enumitem} 
\usepackage[caption=false]{subfig} 

\usepackage[backref]{hyperref} 
\makeatletter
\patchcmd{\@IEEEyesnumber}
  {\stepcounter}
  {\refstepcounter}
  {}{}
\patchcmd{\@@IEEEeqnarray}
  {\stepcounter}
  {\refstepcounter}
  {}{}
\patchcmd{\@@IEEEeqnarraycr}
  {\stepcounter{IEEEsubequation}}
  {\refstepcounter{IEEEsubequation}}
  {}{}
\patchcmd{\@@IEEEeqnarraycr}
  {\stepcounter{IEEEsubequation}}
  {\refstepcounter{IEEEsubequation}}
  {}{}
\patchcmd{\@@IEEEeqnarraycr}
  {\stepcounter{IEEEequation}}
  {\refstepcounter{IEEEequation}}
  {}{}
\patchcmd{\@@IEEEeqnarraycr}
  {\stepcounter{IEEEequation}}
  {\refstepcounter{IEEEequation}}
  {}{}
\makeatother
\usepackage{cleveref} 
\crefname{equation}{eq.}{eqs.}
\crefname{figure}{fig.}{figs.}

\newcommand{\RNum}[1]{\MakeUppercase{\romannumeral#1}} 
\hypersetup{colorlinks=true, linkcolor=black, citecolor=black} 
\begin{document}

\title{Performance of Underwater Wireless Optical Communications in Presents of Cascaded Mixture Exponential-Generalized Gamma Turbulence}

\author{
        Yi~Lou,~\IEEEmembership{Member,~IEEE,}
        Julian~Cheng,~\IEEEmembership{Senior~Member,~IEEE,}
        Donghu~Nie,~\IEEEmembership{Member,~IEEE,}
        and~Gang~Qiao,~\IEEEmembership{Member,~IEEE}
        \thanks{Yi Lou, Donghu Nie, and Gang Qiao are with the Acoustic Science and Technology Laboratory, Harbin Engineering University, Harbin 150001, China, and also with the Key Laboratory of Marine Information Acquisition and Security (Harbin Engineering University), Ministry of Industry and Information Technology, Harbin 150001, China (e-mail: \{louyi, niedonghu, qiaogang\}@hrbeu.edu.cn).}
        \thanks{Julian Cheng is with the School of Engineering, The University of British Columbia, Kelowna, BC, Canada (e-mail: julian.cheng@ubc.ca).}
        \thanks{Manuscript received August 19, 2020; revised August 26, 2020.}
}

\maketitle

\begin{abstract}
        Underwater wireless optical communication is one of the critical technologies for buoy-based high-speed cross-sea surface communication, where the communication nodes are vertically deployed. Due to the vertically inhomogeneous nature of the underwater environment, seawater is usually vertically divided into multiple layers with different parameters that reflect the real environment. In this work, we consider a generalized UWOC channel model that contains $N$ layers. To capture the effects of air bubbles and temperature gradients on channel statistics, we model each layer by mixture Exponential-Generalized Gamma (EGG) distribution. We derive the PDF and CDF of the end-to-end SNR in exact closed-form. Then, unified BER and outage expression using OOK and BPSK are also derived. The performance and behavior of common vertical underwater optical communication scenarios are thoroughly analyzed through the appropriate selection of parameters. All the derived expressions are verified via Monte Carlo simulations.
\end{abstract}


\IEEEpeerreviewmaketitle

\section{Introduction}
\IEEEPARstart{U}{nderwater} wireless optical communications (UWOC) is emerging out as a powerful technology for real-time and ultra-high  underwater applications  whether used alone \cite{zengSurveyUnderwaterOptical2017} or in combination with acoustic counterpart to form so-called acoustic-optical hybrid communications \cite{campagnaroImplementationMultimodalAcousticoptical2016}.

Nevertheless, the performance of UWOC systems is primarily limited by absorption, scattering, and turbulence. Absorption and scattering that depict the energy loss and direction deviation when photons propagate underwater, have been well studied. Turbulence is defined as the rapid fluctuation of the refractive index along the path of the optical field traversing the water medium, caused by the variations in temperature, pressure, salinity, and air bubbles, resulting in signal fading that impairs the UWOC system performance. On the other hand, typical link configurations of UWOC systems include horizontal and vertical configurations. While horizontal configuration has been welled-studied \cite{jamaliPerformanceStudiesUnderwater2017, jamaliMIMOUnderwaterVisible2018, elamassiePerformanceCharacterizationUnderwater2019, celikEndtoEndPerformanceAnalysis2020}, vertical one has not received enough attention.

The difﬁculty of vertical link configuration arises because physics properties of water medium exhibit vertical variability which makes the propagation properties of UWOC channels vertically inhomogeneous as well \cite{smartUnderwaterOpticalCommunications2005, johnsonUnderwaterOpticalWireless2013,anousPerformanceEvaluationNLOS2018}.  Moreover, as the refractive index is depth-dependent, the turbulence characteristics of the vertical UWOC link also vary with depth \cite{nootzQuantificationOpticalTurbulence2016}. Considering the inhomogeneity of turbulence with depth, a layered turbulence model is proposed for the vertical UWOC system, where the turbulence within each layer obeys the same distribution with different parameter values. Both Lognormal \cite{m.elamassiePerformanceCharacterizationVertical2018} and Gamma-Gamma (GG) distributions \cite{elamassieVerticalUnderwaterVisible2020} that are used for modeling atmospheric turbulence, have been adopted for such vertical layered model.

However, the medium that causes stochastic refractive index changes underwater and in the atmosphere is quite different \cite{sahooEstimationChannelCharacteristics2019}, which results that the statistical distributions commonly used in atmospheric turbulence cannot accurately describe underwater turbulence. Recently, using the water tank experiments, a number of models have been proposed to model the turbulence\cite{jamaliStatisticalStudiesFading2018, zediniUnifiedStatisticalChannel2019}. Among them, the Mixture Exponential-Generalized Gamma (EGG) distribution is proposed to characterize the turbulence that takes into account not only the effects of air bubbles, but also the temperature gradients, in both salty and fresh water.

However, to the best of authors’ knowledge, this is the first comprehensive performance analysis of the vertical UWOC system using the layered mixture EGG fading model, which takes into account not only air bubbles and temperature gradients, but also the depth-dependent properties of the refractive index.  We derive the exact closed-form expressions for  ABER, capacity, and outage probability in terms of Fox's H function, which applies to both the types of detection techniques, i.e., heterodyne detection and intensity modulation/direct detection (IM/DD) and serval modulation techniques. Further, we derive corresponding asymptotic expressions with very simple mathematical structures, which is then used to derive the diversity gain. Moreover, the results can be extended and applied to relay-aided UWOC systems.

The rest of this paper is organized as follows. In Section \RNum{2}, the channel and system models are presented. The end-to-end SNR statistics is proposed in \RNum{3}. In Section \RNum{4}, the end-to-end performance metrics are studied. Numerical results are discussed in Section \RNum{5}, followed by the conclusion in Section \RNum{6}.

\section{System and Channel Models}
We consider the same point-to-point vertical UWOC system as that in \cite{elamassieVerticalUnderwaterVisible2020}, where the source node at a depth of $d_0$ vertically communicates with the destination node at a depth of $d_T+d_0$ using IM/DD or heterodyne detection techniques. Considering the non-uniformity of turbulence with depth, we adopt an $N$-layer structure to model the overall UWOC fading. Therefore, the overall turbulence of a $N$-layer channel $\mathcal{I}_N$ can be expressed as
\begin{IEEEeqnarray}{rcl}
        \mathcal{I}_{N}&=&\prod_{n=1}^{N} I_{n}.
\end{IEEEeqnarray}

The normalized fading in each layer is modeled using the mixture EGG distribution for accounting the combined effects of the link loss due to scattering and absorption, and the turbulence due to air bubbles and temperature gradients. The PDF of $I_k$ is given by
\begin{IEEEeqnarray*}{rcl}
        f_{I_i}\left(I_i\right)&=&\frac{c_i \left(1-\omega _i\right)}{I_i \Gamma \left(a_i\right)} \thinspace\text{exp}\left(-\left(\frac{I_i}{b_i}\right)^{c_i}\right) \left(\frac{I_i}{b_i}\right)^{a_ic_i} \\
        &+&\frac{\omega _i }{\lambda _i}\thinspace\text{exp}\left(-\frac{I_i}{\lambda _i}\right), i=1,2,\ldots,N \IEEEyesnumber \label{pdfI}
\end{IEEEeqnarray*}
where $\omega_i$ is the mixture weight of the distribution, $a_i$, $b_i$ and $c_i$ are the parameters related to the exponential distribution, $\lambda_i$ is the parameter related to the exponential distribution, for $i=1,2,\ldots,N$.

To derive the overall turbulence $\mathcal{I}_N$ of a $N$-layer system, we can first aim at the case $N=2$ and derive the PDF of  $\mathcal{I}_2=I_1 I_2$. To facilitate the following derivation, we express the PDF $f_{I_i}\left(I_i\right)$ in \eqref{pdfI} into the form of $H$-function using  \cite[Eq. (2.1.5)]{kilbasHtransformsTheoryApplications2004} and  \cite[Eq. (2.9.1)]{kilbasHtransformsTheoryApplications2004} as
\begin{IEEEeqnarray*}{rcl}
        f_{I_i}\left(I_i\right)&=&\frac{c_i \left(1-\omega _i\right)}{I_i \Gamma \left(a_i\right)}H_{0,1}^{1,0}\!\!\left[\left(\frac{I_i}{b_i}\right)^{c_i}\middle|\!\!\!\begin{array}{c}   \\ \left(a_i,1\right) \\\end{array}\!\!\!\right] \\
        &+&\frac{\omega _i}{\lambda _i}H_{0,1}^{1,0}\!\!\left[\frac{I_i}{\lambda _i}\middle|\!\!\!\begin{array}{c}   \\ (0,1) \\\end{array}\!\!\!\right] \IEEEyesnumber\label{overall1layerpdf}
\end{IEEEeqnarray*}
where $H_{\cdot ,\cdot }^{\cdot ,\cdot }[\cdot|\cdot]$ is the $H$-Function \cite[Eq. (1.2)]{mathaiHFunctionTheoryApplications2010}. Then, using \cite[Eq. (P10.11a.3)]{karbalaygharehChannelModellingPerformance2020}, \cite[Eq. (2.1.5)]{kilbasHtransformsTheoryApplications2004}, and Mellin transformation of two $H$-function \cite[Eq. (2.8.11)]{kilbasHtransformsTheoryApplications2004}, we can express $f_{\mathcal{I}_2}(\mathcal{I}_2)$ as
\begin{IEEEeqnarray*}{rcl}
        f_{\mathcal{I}_2}(\mathcal{I}_2)&=&\int _0^{\infty }\frac{f_{I_1}\left(I_1\right) f_{I_2}\left(\mathcal{I}_2/I_1\right)}{I_1}dI_1\\
        &=&\frac{\left(1-\omega _1\right) \omega _2 }{\mathcal{I}_2 \Gamma \left(a_1\right)}H_{0,2}^{2,0}\!\!\left[\frac{\mathcal{I}_2}{b_1 \lambda _2}\middle|\!\!\!\begin{array}{c}   \\ (1,1),(a_1,\frac{1}{c_1}) \\\end{array}\!\!\!\right] \\
        &+&\frac{\omega _1 \left(1-\omega _2\right) }{\mathcal{I}_2 \Gamma \left(a_2\right)}H_{0,2}^{2,0}\!\!\left[\frac{\mathcal{I}_2}{b_2 \lambda _1}\middle|\!\!\!\begin{array}{c}   \\ (1,1),(a_2,\frac{1}{c_2}) \\\end{array}\!\!\!\right] \\
        &+&\frac{\left(1-\omega _1\right) \left(1-\omega _2\right) }{\mathcal{I}_2 \Gamma \left(a_1\right) \Gamma \left(a_2\right)}H_{0,2}^{2,0}\!\!\left[\frac{\mathcal{I}_2}{b_1b_2}\middle|\!\!\!\begin{array}{c}   \\ \left(a_1,\frac{1}{c_1}\right),(a_2,\frac{1}{c_2}) \\\end{array}\!\!\!\right] \\
        &+&\frac{\omega _1 \omega _2  }{\mathcal{I}_2}H_{0,2}^{2,0}\!\!\left[\frac{\mathcal{I}_2}{\lambda _1 \lambda _2}\middle|\!\!\!\begin{array}{c}   \\ (1,1),(1,1) \\\end{array}\!\!\!\right] \IEEEyesnumber
        \label{overall2layerpdf}
\end{IEEEeqnarray*}

Then, $\mathcal{I}_3=\mathcal{I}_2 I_3$ can be derived using a similar approach. By analogy, noticing the similarity between \ref{overall1layerpdf} and \ref{overall2layerpdf}, we can obtain the overall turbulence $\mathcal{I}_N$ as \eqref{overallpdf}, where $\sum _{i_n\in \{0,1\}}\triangleq\sum _{i_1=0}^1 \sum _{i_2=0}^1 \ldots\sum _{i_N=0}^1$.

The PDF of the instantaneous SNR at the destination node is defined as $\gamma_{N}=\left(\eta \mathcal{I}_{N}\right)^{r} / N_{0}$, where $\eta$ is the effective photoelectric conversion ratio, $N_{0}$ denotes the power of additive white Gaussian noise, and $r$ indicates the types of detection techniques employed, where $r=1$ for heterodyne detection and $r=2$ for IM/DD. $\gamma_{N}$ can be easily derived from \eqref{overallpdf} as \eqref{overallsnrpdf}.

\section{Performance Metrics}
\subsection{Average BER}
A unified average BER expression proposed in \cite{oubeiEfficientWeibullChannel2017},  which is valid for a variety of modulation methods under both heterodyne and IM/DD techniques is given as follows
\begin{IEEEeqnarray*}{rcl}
        P_e=\int _0^{\infty }\underbrace {f_{\gamma _N}(\gamma _N) \Gamma \left(p,\gamma  q_k\right)}_{\Xi}d\gamma\IEEEyesnumber\label{2layerberstep1}
\end{IEEEeqnarray*}
where $p$ and $q_k$ are parameters associated with the modulation and the detection schemes, namely IM/DD and heterodyne detection and is given in \cite[Table \RNum{3}]{zediniUnifiedStatisticalChannel2019}. As with the derivation of \eqref{overall2layerpdf}, we derive the average BER of a 2-layer vertical UWOC system first.

Substituting \eqref{overallsnrpdf} with $N=2$ into \eqref{2layerberstep1} and converting all $H$-functions in  \eqref{2layerberstep1} into the form of line integral \cite[Eq. (2.8.11)]{kilbasHtransformsTheoryApplications2004}, the integrand $\Xi$ can be expressed as
\begin{IEEEeqnarray*}{rcl}
        \Xi&=&-\frac{i \omega _1 \left(1-\omega _2\right) }{2 \pi  \Gamma \left(a_2\right)}\int _{\mathcal{L}}\Gamma (r s+1) \Gamma \left(\frac{r s}{c_2}+a_2\right)\left(\left(b_2 \lambda _1\right)^r \mu _r\right)^s \\
        &\times&\int _0^{\infty }\!\!\!\gamma ^{-s-1} \Gamma \left(p,\gamma  q_k\right)d\gamma ds
        -\frac{i \left(1-\omega _1\right) \omega _2 }{2 \pi  \Gamma \left(a_1\right)}\int _{\mathcal{L}}\Gamma (r s+1)\\
        &\times&\Gamma \left(\frac{r s}{c_1}+a_1\right)\left(\left(b_1 \lambda _2\right)^r \mu _r\right)^s \int _0^{\infty }\gamma ^{-s-1} \Gamma \left(p,\gamma  q_k\right)d\gamma ds \\
        &-&\frac{i \left(1-\omega _1\right) \left(1-\omega _2\right) }{2 \pi  \Gamma \left(a_1\right) \Gamma \left(a_2\right)}\int _{\mathcal{L}}\Gamma\left(\frac{r s}{c_1}+a_1\right) \Gamma \left(\frac{r s}{c_2}+a_2\right) \\
        &\times&\left(\left(b_1 b_2\right)^r \mu _r\right)^s \int _0^{\infty }\gamma^{-s-1} \Gamma \left(p,\gamma  q_k\right)d\gamma ds -\frac{i \omega _1 \omega _2 }{2 \pi }\\
        &\times&\int _{\mathcal{L}}\Gamma (r s+1)^2 \left(\left(\lambda _1 \lambda _2\right)^r \mu _r\right)^s \int_0^{\infty }\gamma ^{-s-1} \Gamma \left(p,\gamma  q_k\right)d\gamma ds\\ \IEEEyesnumber\label{berstep1}
\end{IEEEeqnarray*}

Then, by solving all the integrals with respect to x in \eqref{berstep1}, and using \cite[Eq. (6.451.2)]{i.s.gradshteynTableIntegralsSeries2007},  \cite[Eq. (5.25)]{olverNISTHandbookMathematical2010}, and \eqref{berstep1}, the average BER of the 2-layer vertical UWOC system can be derived in closed-form and extended to the general cases with $N$ layers, which is shown in \eqref{ber} via analogy, where $\mu _r=(\eta  E[I])^r/N_0$.

In the following, we derive the asymptotic average BER expression. For the sake of brevity, we consider the case $N=2$. Noticing that When the transmitting power $\mu _r$ goes to infinity, the values of the variables in all $H$-functions tend to zero. Hence, we use \cite[Eq. (1.8.9)]{kilbasHtransformsTheoryApplications2004} to asymptotically expand all the $H$-functions in \eqref{ber} with $N=2$ near zero and obtain the asymptotic average BER expression in \eqref{asymber}. Moreover, According to the asymptotic BER expression, the diversity order of the 2-layer vertical UWOC system can be easily obtained as $\min \left\{\frac{1}{r},\frac{a_1 c_1}{r},\frac{a_2 c_2}{r}\right\}$.

\begin{figure*}[!bht]
        \begin{IEEEeqnarray*}{rcl}
                f_{\mathcal{I}_N}\left(\mathcal{I}_N\right)=\sum _{i_n\in \{0,1\}} \frac{1}{I}\prod _{n=1}^N \frac{(-1)^{i_n} \omega (n)+i_n}{\Gamma \left(a_n\right)^{i_n}}H_{0,N}^{N,0}\!\!\left[\mathcal{I}_N \prod _{n=1}^N\frac{\left(\frac{\lambda _n}{b_n}\right)^{i_n}}{\lambda _n}\middle|\!\!\!\begin{array}{c}   \\ \left(a_1^{i_1},c_1^{-i_1}\right),\ldots,\left(a_N^{i_N},c_N^{-i_N}\right) \\\end{array}\!\!\!\right]\IEEEyesnumber\label{overallpdf}
        \end{IEEEeqnarray*}
        \hrule
        \begin{IEEEeqnarray*}{rcl}
                f_{\gamma _N}\left(\gamma _N\right)=\sum _{i_n\in \{0,1\}} \frac{1}{\gamma _N}\prod _{n=1}^N \frac{(-1)^{i_n} \omega (n)+i_n}{\Gamma \left(a_n\right)^{i_n}}H_{0,N}^{N,0}\!\!\left[\frac{\gamma_N }{\mu _r} \prod _{n=1}^N \frac{1}{\lambda _n^r}\left(\frac{\lambda _n}{b_n}\right)^{r i_n}\middle|\!\!\!\begin{array}{c}   \\ \left(a_1^{i_1},r c_1^{-i_1}\right),\ldots,\left(a_N^{i_N},r c_N^{-i_N}\right) \\\end{array}\!\!\!\right]\IEEEyesnumber\label{overallsnrpdf}
        \end{IEEEeqnarray*}
        \hrule
        \begin{IEEEeqnarray*}{rcl}
                P_e\!=\!\frac{\delta  }{2 \Gamma (p)}\underset{k=1}{\overset{n}{\sum }}\sum _{i_N\in \{0,1\}}^{ }\!\! \frac{1}{\gamma }\prod _{n=1}^N \frac{(-1)^{i_n}\omega (n)+i_n}{\Gamma \left(a_n\right)^{i_n}}H_{2,N+1}^{N,2}\!\!\left[\frac{1}{q_k \mu _r} \prod _{n=1}^N \frac{\left(\frac{\lambda _n}{b_n}\right)^{ri_n}}{\lambda _n^r}\middle|\!\!\!\begin{array}{c} (1,1),(1-p,1) \\ \left(a_1^{i_1},r c_1^{-i_1}\right),\ldots,\left(a_N^{i_N},r c_N^{-i_N}\right),(0,1) \\\end{array}\!\!\!\right]\IEEEyesnumber\label{ber}
        \end{IEEEeqnarray*}
        \hrule
\end{figure*}

\subsection{Ergodic Capacity}
The overall ergodic capacity  expression is given by
\begin{IEEEeqnarray*}{rcl}
        C=\int _0^{\infty }f_{\gamma_N }(\gamma_N ) \log (\gamma_N  \tau +1)d\gamma_N\IEEEyesnumber \label{capstep1}
\end{IEEEeqnarray*}
where $\tau$ is a constant equal to $\tau =\frac{e}{2\pi }$. We first derive the case $N=2$, we first convert $\log (\gamma_2  \tau +1)$ into the form of $H$-function \cite{zediniUnifiedStatisticalChannel2019} as $H_{2,2}^{1,2}\left[\gamma  \tau \left|\begin{array}{c} (1,1),(1,1) \\ (1,1),(0,1) \\\end{array}\right.\right]$. Substituting the expression of $f_{\gamma_2 }(\gamma_2 )$ obtained from \eqref{overallsnrpdf} by setting $N=2$ into \eqref{capstep1}, we can transform \eqref{capstep1} into the form like \eqref{overall2layerpdf} so that we can solve \eqref{capstep1} in a similar way. Due to the space limit the derivation is omitted. The ergodic capacity expression for a $N$-layer vertical UWOC system is given in \eqref{capacity}.
\begin{IEEEeqnarray*}{rcl}
        C&=&\frac{r \left(1-\omega _1\right) \omega _2 }{\Gamma \left(a_1\right)}H_{1,3}^{3,1}\!\!\left[\frac{\left(b_1 \lambda _2\right)^{-r}}{\tau \mu _r}\middle|\!\!\!\begin{array}{c} (0,1) \\ (0,1),(0,r),(a_1,\frac{r}{c_1}) \\\end{array}\!\!\!\right] \\
        &+&\frac{r \omega _1 \left(1-\omega _2\right) }{\Gamma \left(a_2\right)}H_{1,3}^{3,1}\!\!\left[\frac{\left(b_2 \lambda _1\right)^{-r}}{\tau \mu _r}\middle|\!\!\!\begin{array}{c} (0,1) \\ (0,1),(0,r),(a_2,\frac{r}{c_2}) \\\end{array}\!\!\!\right] \\
        &+&\frac{\left(1-\omega _1\right) \left(1-\omega _2\right) }{\Gamma \left(a_1\right) \Gamma \left(a_2\right)}\\
        &\times&H_{2,4}^{4,1}\!\!\left[\frac{\left(b_1b_2\right)^{-r}}{\tau  \mu _r}\middle|\!\!\!\begin{array}{c} (0,1),(1,1) \\ (0,1),(0,1),(a_1,\frac{r}{c_1}),(a_2,\frac{r}{c_2}) \\\end{array}\!\!\!\right] \\
        &+&r \omega _1 \omega _2 H_{1,3}^{3,1}\!\!\left[\frac{\left(\lambda _1 \lambda _2\right)^{-r}}{\tau  \mu _r}\middle|\!\!\!\begin{array}{c} (0,1) \\ (0,1),(0,r),(1,r) \\\end{array}\!\!\!\right] \IEEEyesnumber
\end{IEEEeqnarray*}

Next, we derive the asymptotic ergodic capacity expression for the $N$-layer system. According to the definition of $H$-function, the path of the integration $\mathcal{L}$ should separates all the poles $b_{j l}=\frac{-b_{j}-l}{\beta_{j}} \quad(j=1, \cdots, m ; l=0,1,2, \cdots)$ to the left and $a_{i k}=\frac{1-a_{i}+k}{\alpha_{i}} \quad(i=1 ; \cdots, n ; k=0,1,2, \cdots)$ to the right. Observing that, for each $H$-function in \eqref{capacity}, the largest pole from left equals to 0, and the smallest pole from right equals to 1, hence the corresponding line integral converges for $0<\mathcal{L}<1$.  Then, the asymptotic expansion of $H$-function at zero can be approximated by evaluating the residues of the line integral at the largest poles from left \cite[Eq. (1.1.2)]{kilbasHtransformsTheoryApplications2004}. After some simplification, the asymptotic expression of ergodic capacity can be expressed in a concise form as \eqref{asymcap}.

\begin{figure*}[!bht]
        \begin{IEEEeqnarray*}{rcl}
                P_{e}^{\text{asy}}&=&\frac{\delta  }{2 \Gamma (p)}\underset{k=1}{\overset{n}{\sum }}-\omega _1 \omega _2\text{  }r^{-1} \Gamma \left(p+\frac{1}{r}\right)\log \left(\frac{\left(\lambda _1 \lambda _2\right)^{-r}}{q_k \mu _r}\right)\text{  }\left(\frac{\left(\lambda _1 \lambda _2\right)^{-r}}{q_k\mu _r}\right)^{\frac{1}{r}}+ \left(1-\omega _1\right) \left(1-\omega _2\right)  \\ &\times&\left(\frac{\Gamma \left(p+\frac{a_1 c_1}{r}\right) \Gamma \left(a_2-\frac{a_1 c_1}{c_2}\right) }{\Gamma \left(a_1+1\right) \Gamma \left(a_2\right)}\left(\frac{\left(b_1b_2\right)^{-r}}{q_k \mu _r}\right)^{\frac{a_1 c_1}{r}}+\frac{\Gamma \left(p+\frac{a_2 c_2}{r}\right) \Gamma \left(a_1-\frac{a_2 c_2}{c_1}\right)}{\Gamma \left(a_1\right) \Gamma \left(a_2+1\right)}\left(\frac{\left(b_1 b_2\right)^{-r}}{q_k \mu _r}\right)^{\frac{a_2 c_2}{r}}\right) \\
                &+&\frac{\omega _1 \left(1-\omega _2\right)}{\Gamma \left(a_2\right)}\left(\!\Gamma \left(p+\frac{1}{r}\right) \Gamma \left(a_2-\frac{1}{c_2}\right) \left(\frac{\left(b_2\lambda _1\right)^{-r}}{q_k \mu _r}\right)^{\frac{1}{r}}-\Gamma \left(-a_2 c_2\right) \Gamma \left(p+\frac{a_2 c_2}{r}\right) c_2 \left(\frac{\left(b_2\lambda _1\right)^{-r}}{q_k \mu _r}\right)^{\frac{a_2 c_2}{r}}\right)  \\
                &+&\frac{\omega _2 \left(1-\omega _1\right)}{\Gamma \left(a_1\right)}\left(\!\Gamma \left(p+\frac{1}{r}\right) \Gamma \left(a_1-\frac{1}{c_1}\right) \left(\frac{\left(b_1\lambda _2\right)^{-r}}{q_k \mu _r}\right)^{\frac{1}{r}}-\Gamma \left(-a_1 c_1\right) \Gamma \left(p+\frac{a_1 c_1}{r}\right) c_1 \left(\frac{\left(b_1\lambda _2\right)^{-r}}{q_k \mu _r}\right)^{\frac{a_1 c_1}{r}}\right) \IEEEyesnumber\label{asymber}
        \end{IEEEeqnarray*}
        \hrule
\end{figure*}

\begin{figure*}[!bht]
        \begin{IEEEeqnarray*}{rcl}
                C=r\!\!\!\sum _{i_N\in \{0,1\}} \prod _{n=1}^N \frac{(-1)^{i_n} \omega (n)+i_n}{\Gamma \left(a_n\right)^{i_n}}H_{2,N+2}^{N+2,1}\!\!\left[\frac{1}{\tau \mu _r}\prod _{n=1}^N \frac{1}{\lambda _n^r}\left(\frac{\lambda _n}{b_n}\right)^{r i_n}\middle|\!\!\!\begin{array}{c} (0,1),(1,r) \\ \left(a_1^{i_1},r c_1^{-i_1}\right),\ldots,\left(a_N^{i_N},r c_N^{-i_N}\right),(0,1),(0,r) \\\end{array}\!\!\!\right]\IEEEyesnumber\label{capacity}
        \end{IEEEeqnarray*}
        \hrule
\end{figure*}

\begin{figure*}[!bht]
        \begin{IEEEeqnarray*}{rcl}
                C^{\text{asy}}=r \left(\underset{k=i}{\overset{n}{\sum }}\frac{\left(1-\omega _i\right) \psi \left(a_i\right)}{c_i}\right)+\log \left(\tau  \mu _r^i\prod _{i=1}^N\lambda _i^{r \omega _i}b_i^{r \left(1-\omega _i\right)}\right)-\gamma  r \underset{k=i}{\overset{n}{\sum }}\omega _i\IEEEyesnumber\label{asymcap}
        \end{IEEEeqnarray*}
        \hrule
\end{figure*}

\section{Numerical and Simulation Results}
Here, we compare the values of the BER and ergodic capacity obtained via Monte Carlo simulations with the derived exact and asymptotic expressions for  the $N$-layered vertical UWOC system in the presence of air bubbles and temperature gradients with both heterodyne and IM/DD schemes to illustrate the correctness of the presented mathematical formulation. The height of each floor $d_i, i=1,2,\ldots,N$ is 1 m. The parameters $\omega_i$, $\lambda_i$, $a_i$, $b_i$ and $c_i$ with $i=1,2,\ldots,N$ (see Lists \RNum{1} and \RNum{2} in \cite{zediniUnifiedStatisticalChannel2019}) obtained from laboratory experiments indicate different levels of air bubbles at constant or gradient temperature, and is shown to have the capacity to model a variety of real ocean environments. For the sake of simplicity, in the following simulations, we use $[\cdot, \cdot]$ to present the values of $[\text{air bubbles level}, \text{temperature gradient}]$.

\begin{figure}[!t]
        \includegraphics[width=.45\textwidth]{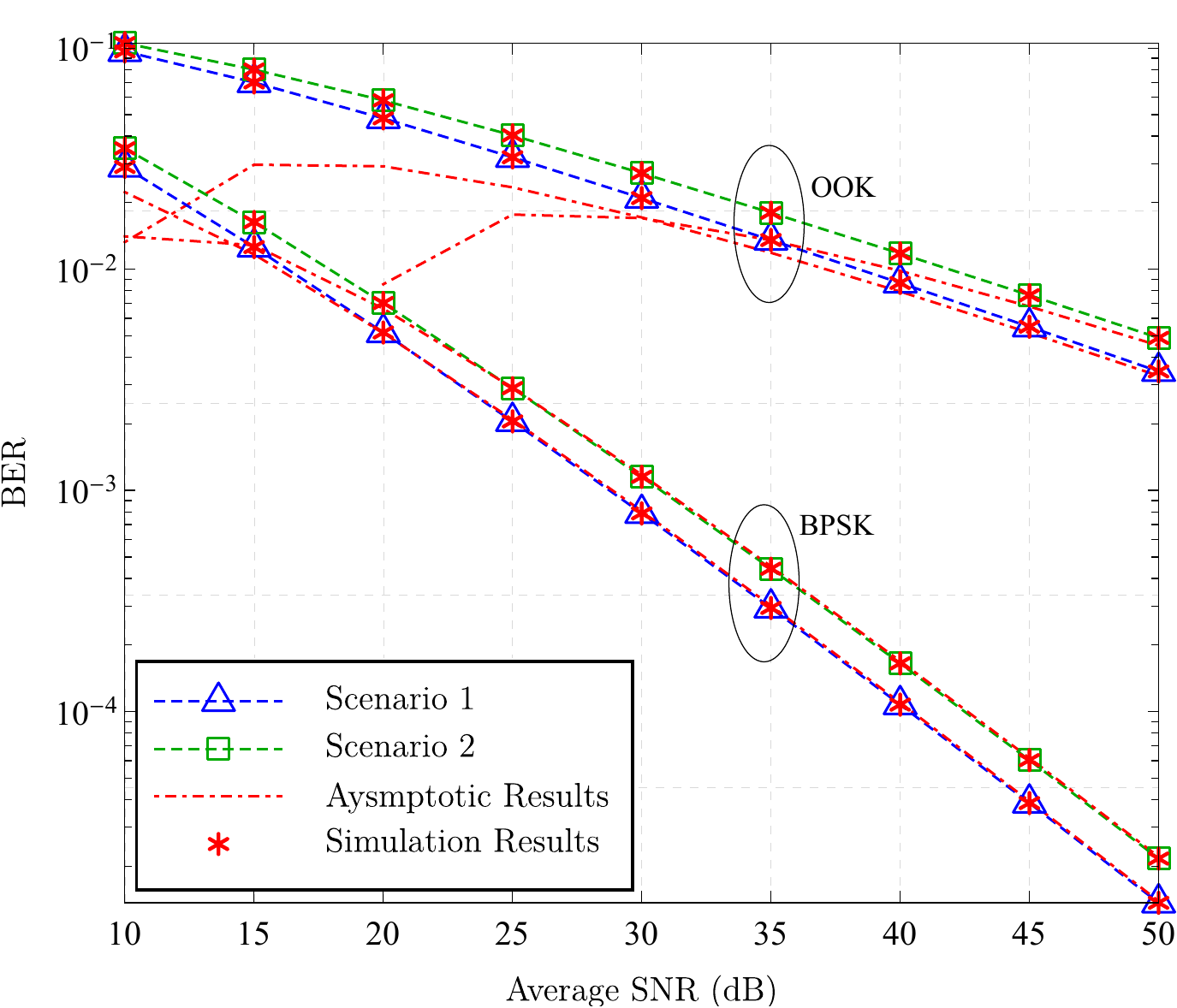}
        \centering
        \caption{Error rate performance of the SWIPT-based DAF with different fading parameters $m$ and $d_0=d_1=d_2=1$, $\eta=0.7$, $\theta=0.5$ and $\alpha=2$. }
        \label{fig: BER2layer}
\end{figure}

In fig. \ref{fig: BER2layer}, we present the BER along with simulation results of a 2-layered 2 m vertical UWOC system under temperature gradient and different levels of air bubbles for both OOK and BPSK schemes. The parameters of the first layer are $[2.4, 0.05]$ in both scenarios. The parameters for the second layer of Scenario 1 and Scenario 2 are  $[2.4, 0.10]$ and  $[2.4, 0.20]$, respectively. We can see from this figure that the analytical and simulation results match perfectly.  Moreover, the level of the air bubbles in the second layer of Scenario 2 is larger than its counterpart of Scenario 1, which causes the overall turbulence of Scenario 2 to be much larger than that of Scenario 1, resulting in stronger turbulence and thus poorer performance in Scenario 1. Also, it can be clearly shown that OOK modulation outperforms BPSK modulation in all SNR regions. Besides, in the high SNR region, the asymptotic and analytical results match very well, which confirms the correctness of our asymptotic expression.

\begin{figure}[!t]
        \includegraphics[width=.45\textwidth]{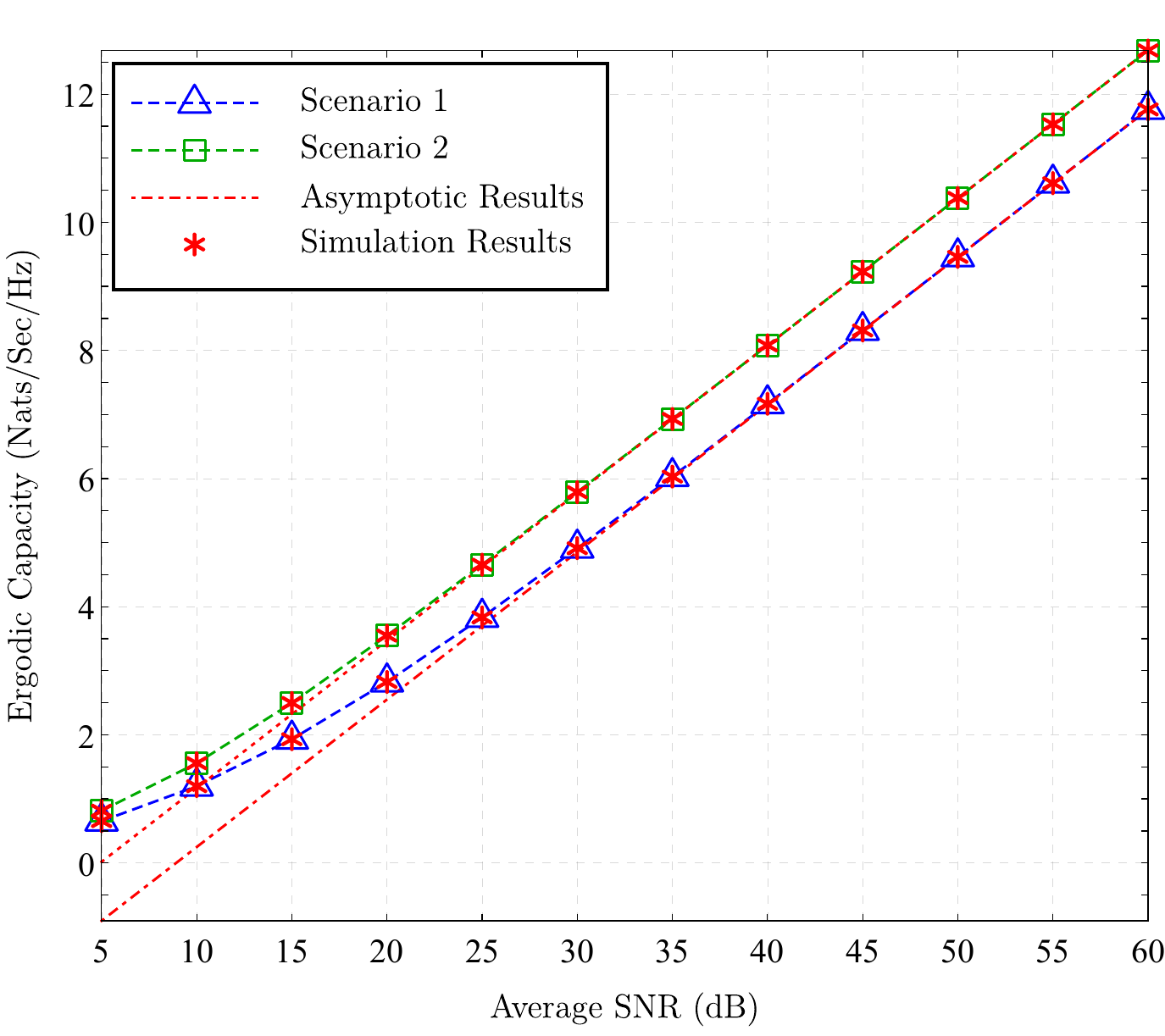}
        \centering
        \caption{Error rate performance of the SWIPT-based DAF with different fading parameters $m$ and $d_0=d_1=d_2=1$, $\eta=0.7$, $\theta=0.5$ and $\alpha=2$. }
        \label{fig: Cap2layer}
\end{figure}

Fig. \ref{fig: Cap2layer} shows the ergodic capacity along with simulation results of a 2-layered 2 m vertical UWOC system under uniform temperature (i.e., the second slot in $[\cdot, \cdot]$ is 0) and different levels of air bubbles for BPSK scheme in salty water. The parameters of the first layer are $[2.4, 0]$ in both scenarios. The parameters for the second layer of Scenario 1 and Scenario 2 are  $[16.5, 0]$ and  $[2.4, 0]$, respectively. We can observe from this figure that as the level of air bubbles increases, turbulence-induced fading increases, resulting in  reduced ergodic capacity. Besides, we can see that the asymptotic expressions of the ergodic capacity not only have simple forms, but also matches the analytical results starting from a small value of SNR for both scenarios. Also, the analytical and simulation results agree very well for all the SNR ranges.

\begin{figure}[!t]
        \includegraphics[width=.45\textwidth]{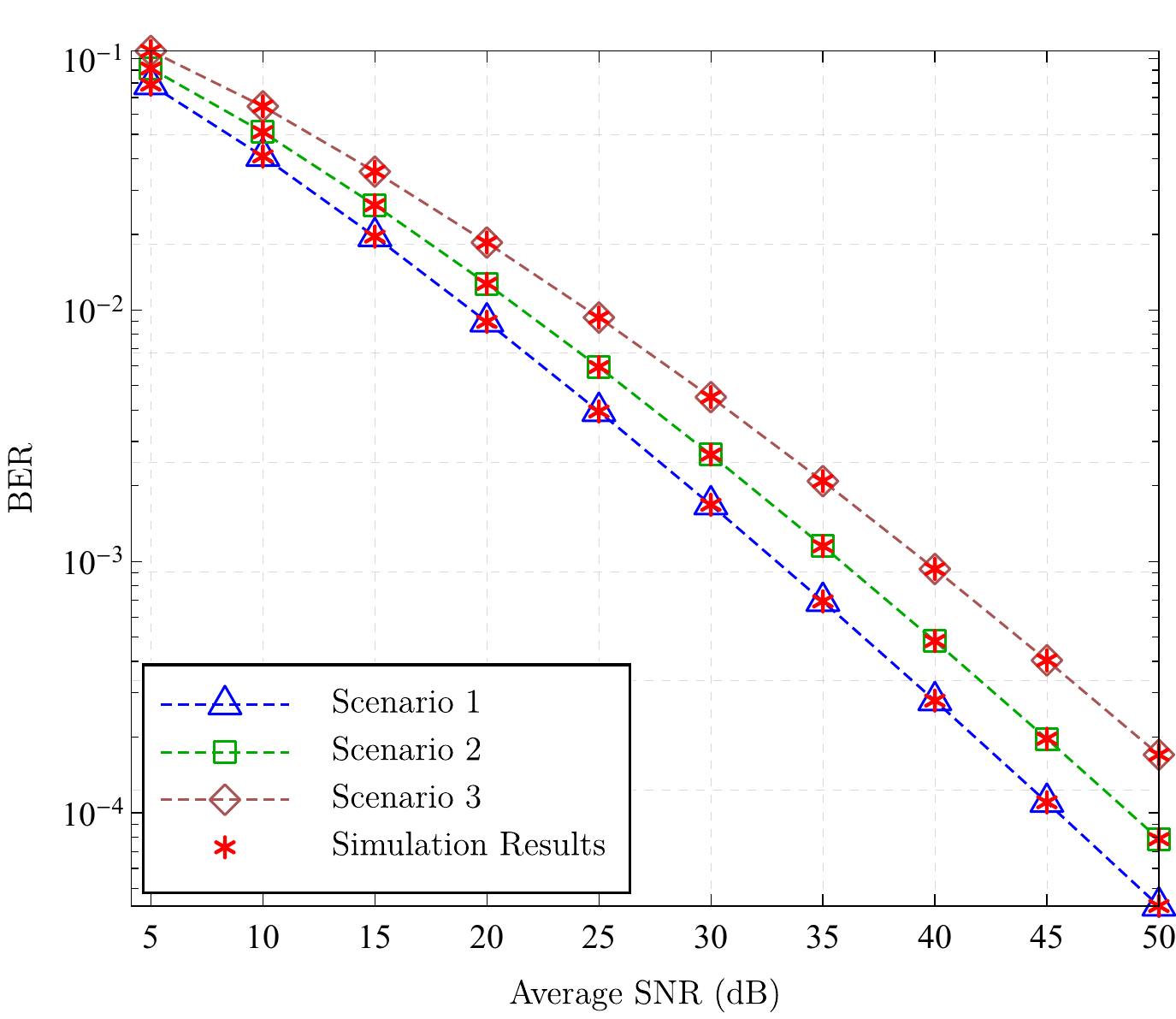}
        \centering
        \caption{Error rate performance of the SWIPT-based DAF with different fading parameters $m$ and $d_0=d_1=d_2=1$, $\eta=0.7$, $\theta=0.5$ and $\alpha=2$. }
        \label{fig: BER3layer}
\end{figure}

The BER along with simulation results of a  3-layered 3 m vertical UWOC system under temperature gradient and same levels of air bubbles for BPSK modulation are shown in fig \ref{fig: BER3layer}. Parameters in each layer for scenario 1 are $[2.4, 0.05]$, $[2.4, 0.05]$, and $[2.4, 0.05]$; for scenario 2 are $[2.4, 0.05]$, $[2.4, 0.10]$, and $[2.4, 0.20]$; for scenario 3 are $[2.4, 0.20]$, $[2.4, 0.20]$, and $[2.4, 0.20]$. We can see from this figure that the analytical results fit well with the simulation results. Moreover, the BER performance of the system is related to the overall turbulence, which is a function of the turbulence in each layer. Specifically, scenario 1 has the smallest temperature gradient in each layer and therefore obtains the best performance. The third set of data has the largest temperature gradient in each layer and therefore the poorest performance. Between these two extremes, scenario 2 obtains the medium performance due to the moderate overall turbulence.

\begin{figure}[!t]
        \includegraphics[width=.45\textwidth]{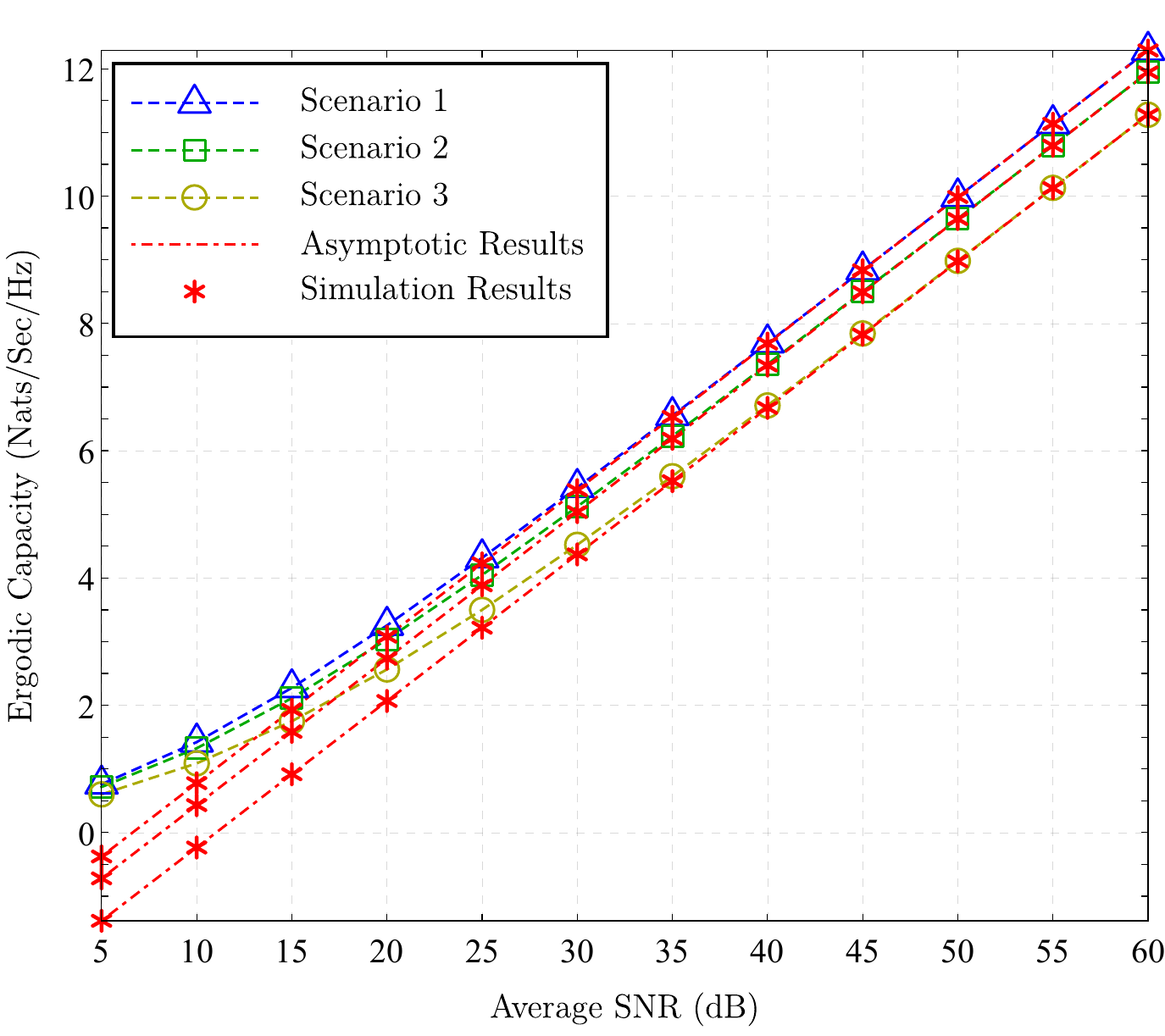}
        \centering
        \caption{Error rate performance of the SWIPT-based DAF with different fading parameters $m$ and $d_0=d_1=d_2=1$, $\eta=0.7$, $\theta=0.5$ and $\alpha=2$. }
        \label{fig: Cap3layer}
\end{figure}

Fig. \ref{fig: Cap3layer} shows the ergodic capacity of a  3-layered 3 m vertical UWOC system under temperature gradient and different levels of air bubbles for BPSK modulation. Parameters in each layer for scenario 1 are $[2.4, 0.05]$, $[2.4, 0.05]$, and $[2.4, 0.05]$; for scenario 2 are $[2.4, 0.05]$, $[2.4, 0.20]$, and $[2.4, 0.20]$; for scenario 3 are $[4.7, 0.05]$, $[4.7, 0.10]$, and $[4.7, 0.10]$. Fig. \ref{fig: Cap3layer} shows that the asymptotic results are tide in high SNR region. Also, the simulation results show a perfect agreement with the analytical results. Moreover,  as shown in this figure, when the temperature gradient or the level of air bubbles decreases, the ergodic capacity of the system increase leading to a system performance improvement.

\bibliographystyle{IEEEtran}
\bibliography{IEEEabrv,mylib.bib}

\begin{thebibliography}{10}
\providecommand{\url}[1]{#1}
\csname url@samestyle\endcsname
\providecommand{\newblock}{\relax}
\providecommand{\bibinfo}[2]{#2}
\providecommand{\BIBentrySTDinterwordspacing}{\spaceskip=0pt\relax}
\providecommand{\BIBentryALTinterwordstretchfactor}{4}
\providecommand{\BIBentryALTinterwordspacing}{\spaceskip=\fontdimen2\font plus
\BIBentryALTinterwordstretchfactor\fontdimen3\font minus
  \fontdimen4\font\relax}
\providecommand{\BIBforeignlanguage}[2]{{%
\expandafter\ifx\csname l@#1\endcsname\relax
\typeout{** WARNING: IEEEtran.bst: No hyphenation pattern has been}%
\typeout{** loaded for the language `#1'. Using the pattern for}%
\typeout{** the default language instead.}%
\else
\language=\csname l@#1\endcsname
\fi
#2}}
\providecommand{\BIBdecl}{\relax}
\BIBdecl

\bibitem{zengSurveyUnderwaterOptical2017}
Z.~Zeng, S.~Fu, H.~Zhang, Y.~Dong, and J.~Cheng,
  ``\BIBforeignlanguage{English}{A {{Survey}} of {{Underwater Optical Wireless
  Communications}}},'' \emph{\BIBforeignlanguage{English}{Ieee Commun Surv
  Tut}}, vol.~19, no.~1, pp. 204--238, 2017.

\bibitem{campagnaroImplementationMultimodalAcousticoptical2016}
F.~Campagnaro, F.~Guerra, P.~Casari, R.~Diamant, and M.~Zorzi, ``Implementation
  of a multi-modal acoustic-optical underwater network protocol stack,'' in
  \emph{Oceans}.\hskip 1em plus 0.5em minus 0.4em\relax {IEEE}, 2016, pp. 1--6.

\bibitem{jamaliPerformanceStudiesUnderwater2017}
M.~V. Jamali, J.~A. Salehi, and F.~Akhoundi,
  ``\BIBforeignlanguage{English}{Performance {{Studies}} of {{Underwater
  Wireless Optical Communication Systems With Spatial Diversity}}: {{MIMO
  Scheme}}},'' \emph{\BIBforeignlanguage{English}{Ieee T Commun}}, vol.~65,
  no.~3, pp. 1176--1192, Mar. 2017.

\bibitem{jamaliMIMOUnderwaterVisible2018}
M.~V. Jamali, P.~Nabavi, and J.~A. Salehi,
  ``\BIBforeignlanguage{English}{{{MIMO Underwater Visible Light
  Communications}}: {{Comprehensive Channel Study}}, {{Performance Analysis}},
  and {{Multiple}}-{{Symbol Detection}}},''
  \emph{\BIBforeignlanguage{English}{Ieee T Veh Technol}}, vol.~67, no.~9, pp.
  8223--8237, Sep. 2018.

\bibitem{elamassiePerformanceCharacterizationUnderwater2019}
M.~Elamassie, F.~Miramirkhani, and M.~Uysal,
  ``\BIBforeignlanguage{English}{Performance {{Characterization}} of
  {{Underwater Visible Light Communication}}},''
  \emph{\BIBforeignlanguage{English}{Ieee T Commun}}, vol.~67, no.~1, pp.
  543--552, Jan. 2019.

\bibitem{celikEndtoEndPerformanceAnalysis2020}
A.~Celik, N.~Saeed, B.~Shihada, T.~Y. {Al-Naffouri}, and M.~S. Alouini,
  ``\BIBforeignlanguage{English}{End-to-{{End Performance Analysis}} of
  {{Underwater Optical Wireless Relaying}} and {{Routing Techniques Under
  Location Uncertainty}}},'' \emph{\BIBforeignlanguage{English}{IEEE Trans.
  Wirel. Commun.}}, vol.~19, no.~2, pp. 1167--1181, Feb. 2020.

\bibitem{smartUnderwaterOpticalCommunications2005}
J.~H. Smart, ``Underwater optical communications systems part 1: Variability of
  water optical parameters,'' in \emph{{{MILCOM}} 2005 - 2005 {{IEEE Military
  Communications Conference}}}, {Atlantic City, NJ}, Oct. 2005, pp. 1140--1146
  Vol. 2.

\bibitem{johnsonUnderwaterOpticalWireless2013}
L.~J. Johnson, R.~J. Green, and M.~S. Leeson, ``Underwater optical wireless
  communications: Depth dependent variations in attenuation,'' \emph{Appl.
  Opt.}, vol.~52, no.~33, pp. 7867--7873, 2013.

\bibitem{anousPerformanceEvaluationNLOS2018}
N.~Anous, M.~Abdallah, M.~Uysal, and K.~Qaraqe,
  ``\BIBforeignlanguage{English}{Performance {{Evaluation}} of {{LOS}} and
  {{NLOS Vertical Inhomogeneous Links}} in {{Underwater Visible Light
  Communications}}},'' \emph{\BIBforeignlanguage{English}{IEEE Access}},
  vol.~6, pp. 22\,408--22\,420, 2018.

\bibitem{nootzQuantificationOpticalTurbulence2016}
G.~Nootz, E.~Jarosz, F.~R. Dalgleish, and W.~Hou, ``Quantification of optical
  turbulence in the ocean and its effects on beam propagation,'' \emph{Appl.
  Opt.}, vol.~55, no.~31, pp. 8813--8820, Nov. 2016.

\bibitem{m.elamassiePerformanceCharacterizationVertical2018}
M.~Elamassie and M.~Uysal, ``Performance {{Characterization}} of {{Vertical
  Underwater VLC Links}} in the {{Presence}} of {{Turbulence}},'' in \emph{2018
  11th {{International Symposium}} on {{Communication Systems}}, {{Networks}}
  \& {{Digital Signal Processing}} ({{CSNDSP}})}, Jul. 2018, pp. 1--6.

\bibitem{elamassieVerticalUnderwaterVisible2020}
------, ``Vertical {{Underwater Visible Light Communication Links}}: {{Channel
  Modeling}} and {{Performance Analysis}},'' \emph{IEEE Trans. Wirel. Commun.},
  pp. 1--1, 2020.

\bibitem{sahooEstimationChannelCharacteristics2019}
R.~Sahoo, S.~K. Sahu, and P.~Shanmugam,
  ``\BIBforeignlanguage{English}{Estimation of the channel characteristics of a
  vertically downward optical wireless communication link in realistic oceanic
  waters},'' \emph{\BIBforeignlanguage{English}{Opt. Laser Technol.}}, vol.
  116, pp. 144--154, Aug. 2019.

\bibitem{jamaliStatisticalStudiesFading2018}
M.~V. Jamali, A.~Mirani, A.~Parsay, B.~Abolhassani, P.~Nabavi, A.~Chizari,
  P.~Khorramshahi, S.~Abdollahramezani, and J.~A. Salehi,
  ``\BIBforeignlanguage{English}{Statistical {{Studies}} of {{Fading}} in
  {{Underwater Wireless Optical Channels}} in the {{Presence}} of {{Air
  Bubble}}, {{Temperature}}, and {{Salinity Random Variations}}},''
  \emph{\BIBforeignlanguage{English}{Ieee T Commun}}, vol.~66, no.~10, pp.
  1--1, Oct. 2018.

\bibitem{zediniUnifiedStatisticalChannel2019}
E.~Zedini, H.~M. Oubei, A.~Kammoun, M.~Hamdi, B.~S. Ooi, and M.~S. Alouini,
  ``\BIBforeignlanguage{English}{Unified {{Statistical Channel Model}} for
  {{Turbulence}}-{{Induced Fading}} in {{Underwater Wireless Optical
  Communication Systems}}},'' \emph{\BIBforeignlanguage{English}{Ieee T
  Commun}}, vol.~67, no.~4, pp. 2893--2907, Apr. 2019.

\bibitem{kilbasHtransformsTheoryApplications2004}
A.~A. Kilbas, \emph{H-Transforms: {{Theory}} and {{Applications}}}, ser.
  Book.\hskip 1em plus 0.5em minus 0.4em\relax {CRC Press}, 2004.

\bibitem{mathaiHFunctionTheoryApplications2010}
A.~M. Mathai, R.~K. Saxena, and H.~J. Haubold, \emph{The {{H}}-{{Function
  Theory}} and {{Applications}}}, ser. Book.\hskip 1em plus 0.5em minus
  0.4em\relax {New York, NY, USA}: {Springer}, 2010.

\bibitem{karbalaygharehChannelModellingPerformance2020}
M.~Karbalayghareh, F.~Miramirkhani, H.~B. Eldeeb, R.~C. Kizilirmak, S.~Q. Sait,
  and M.~Uysal, ``\BIBforeignlanguage{English}{Channel {{Modelling}} and
  {{Performance Limits}} of {{Vehicular Visible Light Communication
  Systems}}},'' \emph{\BIBforeignlanguage{English}{IEEE Trans. Veh. Technol.}},
  vol.~69, no.~7, pp. 6891--6901, Jul. 2020.

\bibitem{oubeiEfficientWeibullChannel2017}
H.~M. Oubei, E.~Zedini, R.~T. ElAfandy, A.~Kammoun, T.~K. Ng, M.~Alouini, and
  B.~S. Ooi, ``Efficient {{Weibull}} channel model for salinity induced
  turbulent underwater wireless optical communications,'' in \emph{2017
  {{Opto}}-{{Electronics}} and {{Communications Conference}} ({{OECC}}) and
  {{Photonics Global Conference}} ({{PGC}})}, {Singapore}, Jul. 2017, pp. 1--2.

\bibitem{i.s.gradshteynTableIntegralsSeries2007}
I.~S. Gradshteyn and I.~M. Ryzhik, \emph{Table of {{Integrals}}, {{Series}},
  and {{Products}}}, 7th~ed., ser. Book.\hskip 1em plus 0.5em minus 0.4em\relax
  {San Diego, CA, USA}: {Academic Press}, 2007.

\bibitem{olverNISTHandbookMathematical2010}
F.~W.~J. Olver, D.~W. Lozier, R.~F. Boisvert, and C.~W. Clark, \emph{{{NIST
  Handbook}} of {{Mathematical Functions}}}, ser. Book.\hskip 1em plus 0.5em
  minus 0.4em\relax {New York, NY, USA}: {Cambridge University Press}, 2010.

\end{thebibliography}
\end{document}